\documentclass[
    ,final            
  ]
  {aipproc}

\layoutstyle{8x11double}

\newcommand{\be}{\begin{eqnarray}}
\newcommand{\ee}{\end{eqnarray}}
\newcommand{\lp}{\left(}
\newcommand{\rp}{\right)}

\begin{document}

\title{The Turbulent Story of X-ray Bursts:\\ Effects of Shear Mixing on Accreting Neutron Stars}

\classification{97.10.Gz; 97.60.Jd; 97.80.-d; 98.70.Qy}
\keywords      {accretion, accretion disks; magnetic fields; neutron stars; X-ray binaries; X-ray bursts}

\author{Anthony L. Piro}{
  address={Astronomy Department and Theoretical Astrophysics Center,
University of California, Berkeley, CA 94720; tpiro@astro.berkeley.edu}
}

\author{Lars Bildsten}{
  address={Kavli Institute for Theoretical Physics, Kohn Hall, University
of California, Santa Barbara, CA 93106; bildsten@kitp.ucsb.edu}
}

\begin{abstract}
During accretion, a neutron star (NS) is spun up as angular momentum is transported through its liquid surface layers. We study the resulting differentially rotating profile, focusing on the impact this has for type I X-ray bursts. The viscous heating is found to be negligible, but turbulent mixing can be activated. Mixing has the greatest impact when the buoyancy at the compositional discontinuity between accreted matter and ashes is overcome. This occurs preferentially at high accretion rates or low spin frequencies and may depend on the ash composition from the previous burst. We then find two new regimes of burning. The first is ignition in a layer containing a mixture of heavier elements with recurrence times as short as $\approx5-30$ minutes, similar to short recurrence time bursts. When mixing is sufficiently strong, a second regime is found where accreted helium mixes deep enough to burn stably, quenching X-ray bursts altogether. The carbon-rich material produced by stable helium burning would be important for triggering and fueling superbursts. 
\end{abstract}


\maketitle


\section{Introduction}
   As neutron stars (NSs) in low mass X-ray binaries accrete material from
their companions they are expected to be spun up by this addition of
angular momentum, possibly becoming millisecond
pulsars \citep{bv91}. This suspicion has received support
by the discovery of accretion driven millisecond pulsars \citep{wv98},
as well as the millisecond oscillations seen during type I X-ray bursts
\citep{cha03}, the unstable ignition of the accumulating fuel
\citep{lvt95,sb06,gal06}. Angular momentum
must be transported into the NS interior if it is to be spun up,
which implies a non-zero,
albeit small, shear throughout the outer liquid parts of the NS.

Such shearing
may lead to viscous heating and chemical mixing at depths far below
the low density boundary layer where the
majority of the shearing occurs.
This has motivated us to assess the importance of angular momentum
transport in NS surface layers. In the following we summarize
the results of our study \citep{pb07}.

\section{Angular Momentum Transport}

Material accreted at a rate $\dot{M}$ reaches the NS surface
with a nearly Keplerian spin frequency of
$\Omega_{\rm K}=(GM/R^3)^{1/2}=1.4\times10^4\ {\rm s^{-1}}\ M_{1.4}^{1/2}R_6^{-3/2}$, where
$M_{1.4}\equiv M/1.4M_\odot$ and $R_6\equiv R/10^6\ {\rm cm}$, which has a kinetic energy
per nucleon of $\approx200\ {\rm MeV\ nuc^{-1}}$. The majority of this energy
is dissipated in a boundary layer of thickness $H_{\rm BL}\ll R$
and never reaches far into the NS surface \citep{is99}.
Nevertheless, angular momentum is added at a rate of
$\dot{M}R^2\Omega_{\rm K}$, so that a torque of this magnitude must be communicated
into the NS. This implies a non-zero shear rate in the interior liquid layers, down to the
solid crust. In the present work we are interested in the shear at the depths where X-ray bursts
ignite, near $\rho\approx10^6\ {\rm g\ cm^{-3}}$, which is well below the boundary layer.

The viscous timescale, $t_{\rm visc}=H^2/\nu$, where $H$ is the pressure scaleheight
and $\nu$ is the viscosity, is always much shorter than the accretion timescale,
$t_{\rm acc}=y/\dot{m}$, where $y$ is the column depth
and $\dot{m}$ is the mass accretion rate per unit area. Therefore we
assume that angular momentum is transported in steady-state. In the limit
$\Omega\ll\Omega_{\rm K}$ and $H\ll R$ this results in \citep{fuj93}
\be
	4\pi R^3\rho\nu q\Omega = \dot{M}R^2\Omega_{\rm K},
	\label{eq:angularmomentum}
\ee
where $q\equiv d\log\Omega/d\log r$ is the shear.
This equation shows that $q>0$ when angular momentum is transported
inward. In general, we find that the shearing is rather small ($q\ll1$)
at the depths of interest. Nevertheless $q$ is large enough to
activate instabilities that help to transport angular momentum,
as well as mix material.

\section{Turbulent Mixing}

We use equation (\ref{eq:angularmomentum}) to evaluate the shear for various turbulent
viscosities and find that purely hydrodynamic instabilities
\citep{fuj93} are insufficient to prevent strong shearing of magnetic fields.
This leads to generation of the Tayler-Spruit dynamo \citep{spr02}, where toroidal
field growth is balanced by Tayler instabilities to create a steady-state magnetic field,
which provides a torque on the shearing layers.
The result is nearly uniform rotation and little viscous heating,
too little to affect either X-ray bursts or superbursts, thermonuclear
ignition of ashes from previous X-ray bursts \citep{cb01,sb02}.
Turbulent mixing is found to be non-negligible and in some cases may mix fresh material
with the ashes of previous bursts, therefore we focus on this for the majority of our study.

\subsection{The Buoyancy Barrier}

A key point for whether mixing can occur is whether the strong buoyancy due to the larger density of
the ashes below can be overcome. Analytic analysis shows that for this to occur the accretion rate must
exceed
\be
	\dot{m}_{\rm crit,1}\approx5\times10^{-2}\ \dot{m}_{\rm Edd}\ \alpha_{\rm TS}^{-3}
		\lp\frac{\Omega}{0.1\Omega_{\rm K}}\rp^3\lp\frac{\Delta\ln\mu}{0.44}\rp,
	\label{eq:mdotcrit1}
\ee
where $\dot{m}_{\rm Edd}= 1.5\times10^5\ {\rm g\ cm^{-2}\ s^{-1}}R_6^{-1}$,
is the local Eddington limit for helium-rich accretion, $\alpha_{\rm TS}$ is a dimensionless
constant (of order unity) that adjusts the strength of mixing, and $\Delta\ln\mu$ is the
fractional change in mean molecular weight across the boundary ($\approx0.44$ for
helium-rich material above an iron-rich ocean). This critical $\dot{m}$
depends on the composition of the ashes from previous bursts (which sets $\Delta\ln\mu$) so
that the strength of mixing could very well change from burst to burst. 

\subsection{Mixed Unstable Ignition}

\begin{figure}
  \includegraphics[height=.3\textheight]{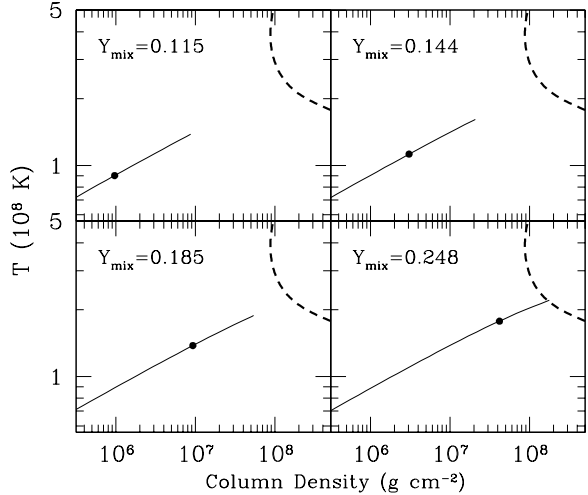}
  \caption{The four panels show how the fully mixed accumulating layer evolves
in time until it reaches conditions necessary for unstable ignition. In each
panel, the column of helium that has been accreted, $y_{\rm acc}$, is denoted by a filled circle.
 Mixing takes place down to the column reached by the thin solid line, $y_{\rm mix}$.
 The mixed helium fraction,
$Y_{\rm mix}=y_{\rm acc}/y_{\rm mix}$ is displayed in the upper left-hand corner of each panel. Helium ignition
curves are shown as a thick dashed line.}
\label{fig:mixed_ignition}
\end{figure}

We next assess the affect on the burst properties when mixing occurs.
Figure \ref{fig:mixed_ignition}
shows an example calculation of material accumulating and mixing on the NS surface causing
premature ignition of the helium, for the parameters $\dot{m}=0.1\dot{m}_{\rm Edd}$
and $\Omega=0.1\Omega_{\rm K}$. The shear profile is evaluated using equation
(\ref{eq:angularmomentum}), and we assume that mixing occurs
completely down to a depth $y_{\rm mix}$
at which $t_{\rm acc}=t_{\rm mix}$ (the right end of each thin solid
line), where $t_{\rm mix}=H^2/D$
and $D$ is the turbulent mixing diffusivity of the Tayler-Spruit dynamo \citep{spr02}. This is
significantly deeper than the amount of material accreted, $y_{\rm acc}$ (shown by the filled circles).
This means that helium is significantly depleted at the time of ignition by an amount
$Y_{\rm mix}=y_{\rm acc}/y_{\rm mix}$, where we use a helium-rich composition for the accreted
material.

\begin{figure}
  \includegraphics[height=.29\textheight]{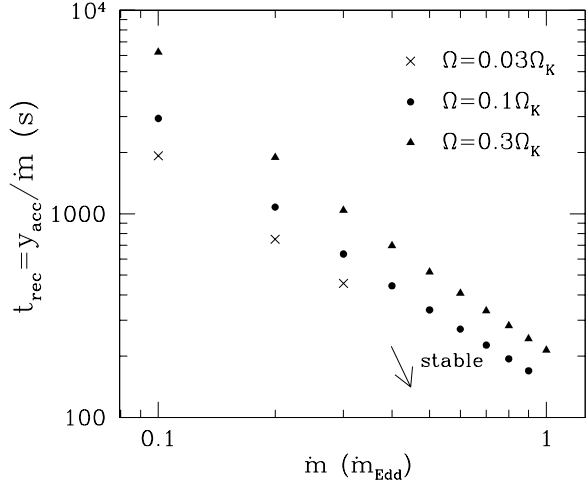}
\caption{The recurrence time for mixed-ignition models as a function
of $\dot{m}$. The symbols denoted different spins, as shown in the key.
Models that are at sufficiently high $\dot{m}$ or low $\Omega$ do not ignite
unstably, and thus are not plotted.}
\label{fig:trec}
\end{figure}

In Figure \ref{fig:trec} we plot the recurrence times for mixed bursts (once again assuming
that eq. [2] is satisfied). These timescales are similar to those found for short recurrence
time bursts \citep{boi07}. To explain the energetics of these bursts (i.e. their low $\alpha$-value
\citep{boi07}) still
requires incomplete burning from the previous burst. \emph{This is in fact preferential for strong
mixing since incomplete burning also leads to smaller compositional gradients.}
If the mixing is too strong we instead find stable helium burning (noted in Fig. \ref{fig:trec}).
This is analytically estimated to occur at
\be
	\dot{m}_{\rm crit,2}\approx\dot{m}_{\rm Edd}\ \alpha_{\rm TS}^{-0.83}
		\lp\frac{\Omega}{\Omega_{\rm K}}\rp^{0.62}.
	\label{eq:mdotcrit2}
\ee
Above this $\dot{m}$ we expect X-ray bursts to cease altogether.

\section{Conclusions}
\label{sec:theend}

\begin{figure}
  \includegraphics[height=.3\textheight]{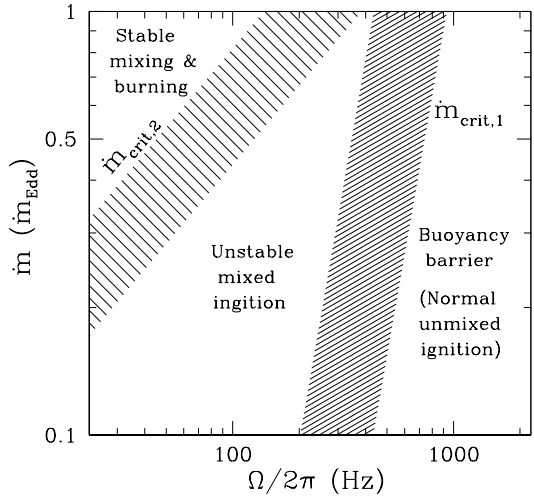}
\caption{The three regimes of burning found for models including mixing. The boundaries between each regime are shaded to emphasize uncertainty in the strength of mixing ($\alpha_{\rm TS}=0.7-1.5$). The light shaded region divides between stable and unstable mixed burning
($\dot{m}_{\rm crit,2}$, eq [\ref{eq:mdotcrit2}]). The heavy shaded region corresponds to the compositional barrier, which prevents mixing ($\dot{m}_{\rm crit,1}$, eq [\ref{eq:mdotcrit1}]). Its position can move significantly depending on the composition of material from the previous burst.}
\label{fig:phase}
\end{figure}

Turbulent mixing is sufficiently large to have important consequences
for X-ray bursts. We constructed simple models, both analytic and numerical,
to explore mixing for pure helium accretion \citep{pb07}. From these models we can make
a few conclusions that are likely general enough to apply to most viscous mechanisms.
As a guide, we show the different burning regimes
we find in Figure \ref{fig:phase}. These can be summarized as follows:
\begin{itemize}
\item Mixing has trouble overcoming the buoyancy barrier at chemical discontinuities. Incomplete burning results in small compositional gradients so that mixing is important in subsequent bursts.
\item Mixing is strongest at large $\dot{m}$ (when angular momentum is
being added at greater rates) and small $\Omega$ (which gives a larger
relative angular momentum between the NS and accreted material).
\item Mixing can lead to two new effects. First, the layer may ignite, but in a mixed environment with a shorter recurrence time. Second, for strong mixing accreted helium can mix and burn in steady-state, quenching bursts. Both regimes have observed analogs, namely the short recurrence time bursts
\citep{boi07} and the stabilization of bursting seen near one-tenth the Eddington rate \citep{cor03}.
\end{itemize}
As a future extension of this work, realistic ash compositions from previous bursts should
be incorporated into simulations, so that one can test whether mixing is a suitable explanation
for short recurrence time bursts. Mixed ignition can be studied using current
numerical experiments \citep{woo04} by just artificially accreting
fuel with a mixed composition of $Y_{\rm mix}\approx0.1-0.6$. These calculations are simplified by
our conclusion that shearing and heating can be ignored, at least for initial
studies. Another important
extension would be to use different compositions for the accreted material.
The current models assume it is pure-helium.
Including hydrogen would allow comparison to a wider range of systems. Finally, the carbon-rich
ashes from periodic stable burning may be important for explaining the recurrence times
of superbursts \citep{cb01,sb02,cum06}, which also deserves further investigation.


\begin{theacknowledgments}
We thank Henk Spruit for helpful discussions.
This work was supported by the National Science Foundation
under grants PHY 99-07949 and AST 02-05956.
\end{theacknowledgments}


\bibliographystyle{aipproc}   


\bibliography{sample}


\end{document}